# Exploring Spreadsheet Use and Practices in a Technologically Constrained Setting

Khwima Mckinley Mkamanga
*kmkamangah@gmail.com*
Simon Thorne, Cardiff Metropolitan University
*SThorne@cardiffmet.ac.uk*

**ABSTRACT**

*This paper explores the impacts of spreadsheets on business operations in a water utility parastatal in Malawi, Sub-Saharan Africa. The organisation is a typical example of a semi-government organisation operating in a technologically underdeveloped country. The study focused on spreadsheet scope of use and life cycle as well as organisation policy and governance. The results will help define future spreadsheet usage by influencing new approaches for managing potential risks associated with spreadsheets in the organization. Generally, findings indicate that the proliferation of spreadsheets in the organization has provided an enabling environment for business automation. The paper also highlights management, technological and human factor issues contributing to high risks associated with the pervasive spreadsheet use. The conclusions drawn from the research confirms that there is ample room for improvement in many areas such as implementation of comprehensive policies and regulations governing spreadsheet development processes and adoption.*

## 1.0 Introduction

Spreadsheet practices presented in this paper are based on a study conducted in a water utility public organisation operating in a least developed country. The activities of the organisation include management of water treatment and distribution process, billing and cash collection processes as well as managing and implementing national water projects which are largely financed by the Government and International Development partners. This paper is based on the data that was generated through a detailed questionnaire, unstructured interviews, spreadsheet audit using software tools and review of existing spreadsheet artefacts stored on the organisation's computers and network.

### 1.1 Aim and Objectives

This study was aimed at examining the scope of spreadsheet use, practices, and risks associated with spreadsheet use in a technologically constrained environment.

### 1.2 Research Questions

The research has set out general research questions to be addressed during the study as follows:

a) What is the level of importance of spreadsheets?
b) What processes are involved (e.g. planning, testing, documentation) when developing spreadsheet models?

c) What measures are put in place to protect spreadsheets from abuse?
d) How are critical spreadsheets archived or backed up?
e) Are there policies or regulations that govern the use of spreadsheets?

## 2.0 Method

The organisation has three business units referred to as departments – Finance, Operations and Human Resources and Administration. The study investigated all the three business units in order to gain an overall insight into spreadsheet practices and activities. In addition, the absence of comprehensive Management Information Systems (MIS) that would cater for all business processes in the organisation suggests the likelihood of finding high concentrations of spreadsheet users in all business units of the organisation.

### 2.1 Case Organisation

This paper focuses on an in-depth investigation of spreadsheet practices and risks in a water utility organisation. The organisation has over 500 employees and operates in 21 offices and water reticulation plants in Malawi. Survey data was sought from different sources within the organisation through questionnaires, semi-structured interviews and preliminary review of existing files. Apart from that, spreadsheet auditing tools (XLTest and Operis Analysis Kit) were used to generate data by auditing a critical spreadsheet model that was selected from the operations department. Currently, poor ICT infrastructure in Malawi generally affects the implementation and reliability of technology services in business operations. This explains why the study discovered that many times employees are forced to capture various transactions manually into spreadsheets despite the fact that the organisation adopted an Enterprise Resource Planning (ERP) solution.

### 2.2 Research process

The research involved carrying out three key processes. First, a preliminary review of existing spreadsheets and documents was conducted along with the survey to collect data pertaining to three aspects of spreadsheets - general perception, typical spreadsheet life cycle and organization policies. The survey was based on the questionnaire that was used during the Tuck Spreadsheet Engineering Research Project (SERP) that was conducted by a team of researchers from Tuck School of Business at Dartmouth (Baker et al. 2006). The main focus was to understand levels of spreadsheet use; approaches to spreadsheet creation; training acquired and availability; organisation standards and policies and spreadsheet risk awareness. Second, the researcher conducted semi-structured interviews with key informants to gain a deeper understanding on how spreadsheets were developed and controlled in the organisation. This activity was undertaken to augment data collected through questionnaire responses. As pointed out by Bell (1993), 'semi-structured interviews centred around the topic area help to produce a wealth of valuable data.' The third process of the study involved auditing spreadsheets using application tools. The tools effectively indexed potential risks from a sample of critical spreadsheets used in the organisation.

## 3.0 Results

This section considers data that was produced from the questionnaires administered in the organisation. Interpreted results will be presented to show any trends and relationships which can lead to spreadsheet risks.

### 3.1 Preliminary Review

The review process discovered huge volumes of duplicate spreadsheets stored on computers across the organisation. High traffic of spreadsheets circulated via emails is linked to over-dependence on spreadsheet models in the organisation as well as a lack of a reliable and comprehensive MIS solution. In addition, the technology infrastructure both at organizational and national levels is heavily lacking leading to a high prevalence of poor spreadsheet practices. Currently, business processes in the organisation are largely not computerised hence spreadsheets are adopted to offset the gap. As a result, the company is paying to store a huge volume of spreadsheets that probably are no longer relevant and could even present privacy challenges to the organisation.

### 3.2 Questionnaire Responses

This section presents results from questionnaire responses. The researcher distributed 80 questionnaires to a sample of employees in all the 21 offices and a total of 71 questionnaires were completed representing 89% response rate from subjects who are the employees from the Finance, Operations and HRA departments

#### 3.2.1 Sample Characteristics

The last part of the questionnaire was aimed to establish the characteristics of the respondents. Results show that 42% of respondents work in operations, 38% in finance and 20% indicated administration as shown in figure 4.1. These responses were a collection of a distinctive sample of people who use spreadsheets. Therefore, this implies that the sample is heterogeneous in the sense that different categories of people with varied professional backgrounds were involved in the survey, the sample included executives with extensive work experience and trained to manage demanding business processes while others have little prior experience in using spreadsheets. As already mentioned above, the absence of a comprehensive MIS in the organization contributes to over-dependence on spreadsheets especially in the operations department. In terms of academic qualifications, 63% of the sample are graduates, 18% hold postgraduate degrees which also tallies with the number of employees holding other non-degree professional qualifications. Results further show that 2% of respondents are executives, 27% are managers 68% were non-managers and casual employees accounted for only 2%.

**Figure 1:** Respondents based on departments in the organisation

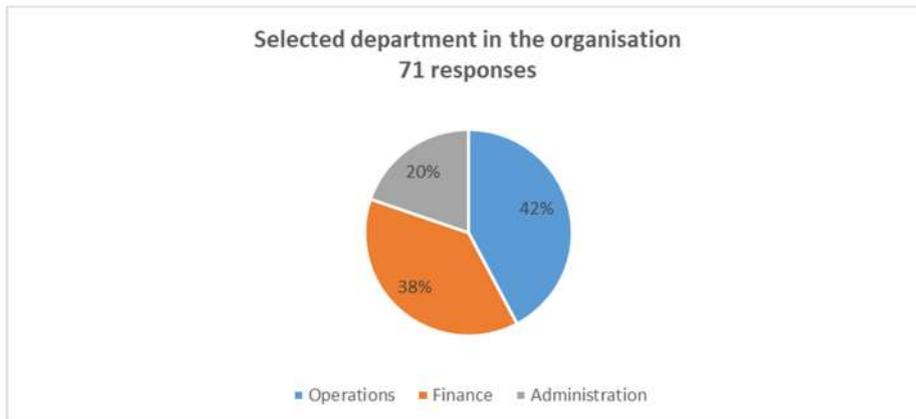

### 3.2.2 Spreadsheet Use

In this section, data was sought relating to users' perceptions and information about spreadsheet application areas in the organisation. Firstly, the word "Spreadsheet" was seemingly unfamiliar to some members of the sample as they are more accustomed to "Excel or Excel sheets". This helps explain why 100% of the sample indicated Microsoft Excel as a spreadsheet application used in the organisation. The other spreadsheet applications like Google sheets, OpenOffice and LibreOffice are not known in the organisation. Survey results further show that spreadsheets are frequently used on a weekly and monthly basis to compute different tasks such as data analysis, tracking of data, recording lists, defining trends and projections as well as evaluating alternatives. The last question under this section probed the level of spreadsheets' importance in the organisation. Figure 2 has been used to summarise results. As shown, over 70% of the staff think spreadsheets are critical to performing the roles of their jobs in the organisation. These results help to highlight the need for ensuring that managerial, technological and human factor issues provide a good spreadsheeting environment. Thus, spreadsheeting requires serious attention from all stakeholders in the form of best practices, which is very broad in scope and is deeply explored in the subsequent sections.

**Figure 2:** Level of spreadsheet importance

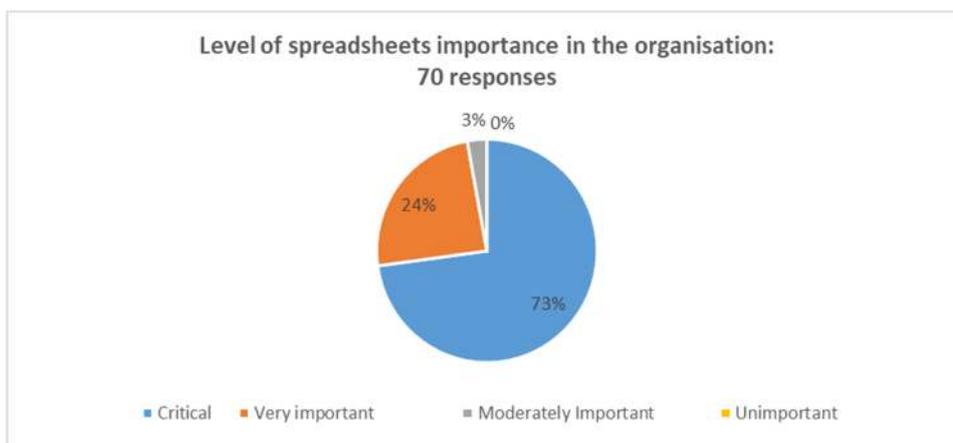

### 3.2.3 Spreadsheet Design

As the preliminary review established that there is a high dependence on spreadsheets in the organisation, it is therefore crucial to understand how models are developed since the design phase forms a critical part of the life cycle of typical spreadsheets. According to studies that were conducted by (Teo and Lee-Partridge, 2001; Teo and Tan, 1997), poorly designed spreadsheets tend to pose challenges when debugging and detecting errors. This is reflected by the results presented in tables 4.1 and 4.2 in which over 80% of spreadsheets were not built from scratch, supported by the results showing that 66% of the respondents indicated never as well as 24% who think they rarely design models from the scratch. Contrary to the best design practices, only 6% of employees in the organisation sketch their spreadsheets prior to design. Baker et al. (2006) argue that sketching spreadsheets help in reducing errors that arise due to the rework and modification processes during later stages of development. These practices can lead to risks of inheriting errors in cases of design flaws from the 37% of employees who initiate the spreadsheet design process by simply entering data and formulas directly into the computer.

| *Table 1: Designing spreadsheets from scratch* | *Responses* | *Percentage* |
|---|---|---|
| *Usually* | *47* | *66%* |
| *Never* | *17* | *24%* |
| *Sometimes* | *5* | *7%* |
| *Always* | *2* | *3%* |

| *Table 2: Initial step while designing spreadsheets* | *Responses* | *Percentage* |
|---|---|---|
| *Borrow a design* | *38* | *57%* |
| *Enter data/formulas directly* | *25* | *37%* |
| *Sketch on paper* | *4* | *6%* |
| *Write relational algebra* | *0* | *0%* |

The preliminary review exercise revealed the existence of complex spreadsheet models, especially in the Finance and Operations departments. Whilst the findings of the study on 106 spreadsheet developers in Australia by (Hall,1996) did not conform to acceptable practices, the situation in this study is by far worse, as only about 16% of spreadsheets have modular designs as shown in table 3.

| *Table 3: Separating into discrete modules:* | *Responses* | *Percentage* |
|---|---|---|
| *Usually* | *47* | *66%* |
| *Never* | *13* | *18%* |
| *Sometimes* | *7* | *10%* |
| *Always* | *4* | *6%* |

| *Table 4: Sheets are used to separating inputs and formulas?* | *Responses* | *Percentage* |
|---|---|---|
| *Usually* | *38* | *56%* |
| *Never* | *14* | *21%* |
| *Sometimes* | *13* | *19%* |
| *Always* | *3* | *4%* |

Survey results presented in Tables 3 and 4 suggest the existence of significant room for improvement to curtail challenges relating to possible code inspection or debugging activities (Baker et al. 2006). The last question in this section was aimed at verifying how formal and seriously a typical spreadsheet life cycle is considered. It was further discovered that 99% of the employees do not adhere to standard system development methodologies like the Systems Development Life Cycle (SDLC) when developing spreadsheets.

### 3.2.4 Testing and Documenting

Kimberland (2004) attests that software testing in conventional programming consumes about 40% of the development time. However, studies by Panko (2005, 2006b) have demonstrated that error rates in spreadsheets and conventional programming spreadsheeting are analogous despite the fact that testing among spreadsheet modelers is extremely rare. In this study, findings reveal that spreadsheet testing is also not commonly practiced as denoted by 58% and 24% of respondents who selected "never" and "rarely" respectively whilst 77% of a small group of employees who test their models indicated that they use common sense as shown in figure 4.3 and table 4.5. These results simply confirms that most employees in the organization are overconfident about the quality of their models, contributing to a high possibility of errors (Panko and Halverson, 1997; Brown and Gould, 1987).

**Figure 3:** Spreadsheet Testing

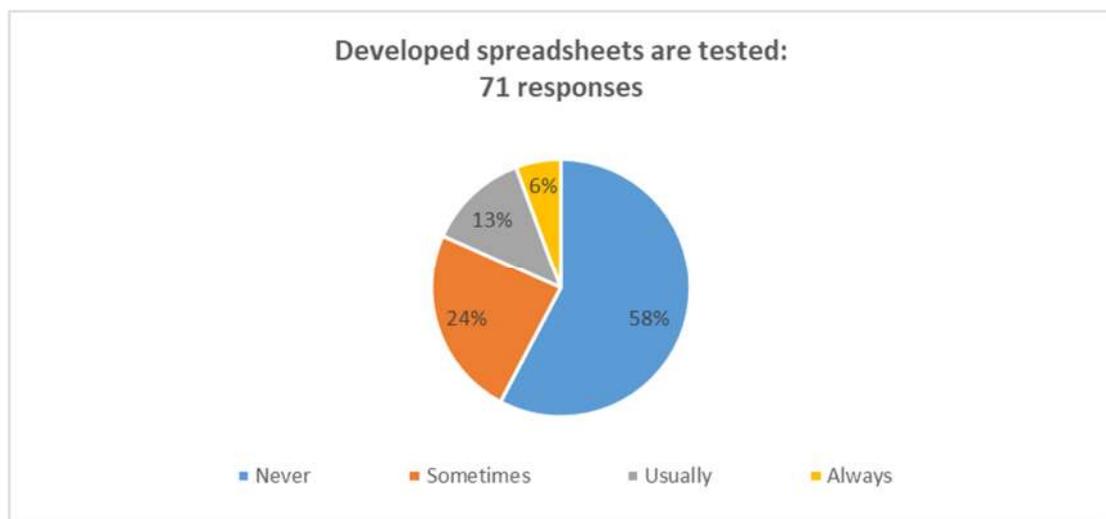

| Table 5: Methods for testing spreadsheets | Responses | Percentage |
|---|---|---|
| Use common sense | 23 | 77% |
| Calculator to check cells | 12 | 40% |
| Display formulae | 11 | 37% |
| Testing the performance for credibility | 7 | 23% |
| Error checking option | 7 | 23% |
| Examine formulas independently | 6 | 20% |
| Toolbar for formula auditing | 4 | 13% |
| Test extreme cases | 2 | 7% |
| Go To - Special | 0 | 0% |

Apparently, documenting spreadsheets in the organisation is not taken seriously as evidenced by a thorough review of sampled files which did not show documentation traces of any type. The picture was more depressing after results revealed that only 12% of employees specified that they "usually" or "always" document their model as summarised in Figure 4. Additionally, Table 4.6 displays that 85% of the sample that performs some documentation activities only spend less than 5% of the time devoted to spreadsheeting.

**Figure 4:** Spreadsheet Documentation

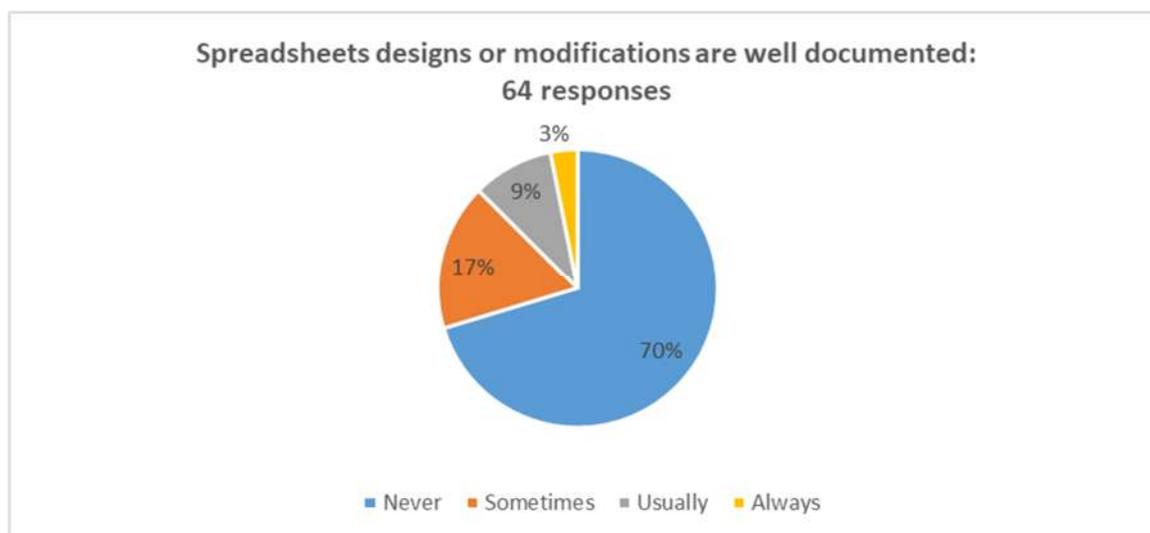

| *Table 6: Time devoted to spreadsheet documentation* | *Responses* | *Percentage* |
|---|---|---|
| *1 - 5%* | *23* | *85%* |
| *6 - 15%* | *3* | *11%* |
| *Over 15%* | *1* | *4%* |

Finally, this theme also probed techniques that are employed when documenting spreadsheets. Statistics in Table 7 support the argument that documentation is rare in spreadsheets (Panko, 1998). They reveal that employees simply use cell comments or text in their spreadsheets as approaches to document their spreadsheet models.

| *Table 7: Techniques for documenting spreadsheets* | *Responses* | *Percentage* |
|---|---|---|
| *Cell comments* | *18* | *90%* |
| *Text in spreadsheets* | *13* | *65%* |
| *Separate document* | *6* | *30%* |
| *Separate sheet* | *4* | *20%* |
| *Others* | *0* | *0%* |

### 3.2.5  Sharing, Protecting and Modifying

Responses to the question about the approaches that are adopted to ensure protection and version control of shared spreadsheets, 92% of the sample selected "no protection" and spreadsheet versions are commonly tracked by saving with date or user. In spreadsheeting,

EuSpRIG 2021 Proceedings. ISBN: 978-1-905404-57-5 Copyright © 2021, European Spreadsheet
Risks Interest Group (www.eusprig.org) & the Author(s)
Page 7 of 25

version control enables tracking of the iterative changes made during the life cycle. This allows experimentation of new concepts with the option of reverting to a specific past version used to generate particular results. The responses presented in Tables 8 and 9 below demonstrate that 75% of employees circulate the whole spreadsheet, usually on a weekly or monthly basis. Such practices in this organisation can increase the risks of secure data landing in the wrong hands. The results in the two tables expose gaps between current practices in the organization and best spreadsheet practices as advocated by (Alexander and Sheedy, 2005). Although (Panko, 2002) argues that errors are common spreadsheet risks, it is not clear whether sharing has a considerable impact on this risk.

| *Table 8: The manner on how spreadsheets are shared* | *Responses* | *Percentage* |
|---|---|---|
| *Whole spreadsheet* | *39* | *75%* |
| *Summarised results* | *11* | *21%* |
| *Parts of the spreadsheet* | *10* | *19%* |
| *Rarely shared* | *4* | *8%* |

| *Table 9: Sharing frequency* | *Responses* | *Percentage* |
|---|---|---|
| *Monthly* | *28* | *54%* |
| *Weekly* | *22* | *42%* |
| *Daily* | *8* | *15%* |
| *Annually* | *3* | *6%* |

Baker et al. (2006) contends that modifying spreadsheets requires the process of revising key stages of development like design, test and documentation. In an ideal situation, these processes are supposed to be done by an expert in the field and not necessarily the original developer. Table 10 results show that 52% of users are allowed to modify models, raising questions on whether users' expertise matches spreadsheet modelers which can lead to devastating effects of spreadsheet errors.

| *Table 10: Modifier of spreadsheets* | *Responses* | *Percentage* |
|---|---|---|
| *Original developer* | *31* | *44%* |
| *New developer* | *3* | *4%* |
| *Any users* | *37* | *52%* |

### 3.2.6 Archiving Spreadsheets

The researcher was fully aware that technologies are very lacking in the organisation to support the fast-growing demand for business automation. However, it was worthy to get an account of peoples' sentiments on how huge volumes of spreadsheets are preserved. Clearly, two major approaches are adopted to store spreadsheets as presented in Figure 5 below. The majority (61%) depends on storage available on the C:\ directory of their local Computer whilst 24% of the sample resort to removable flash disks. These practices are common due to lack of central repositories like SharePoint or file server implementations.

**Figure 5:** Spreadsheet Back-up approaches

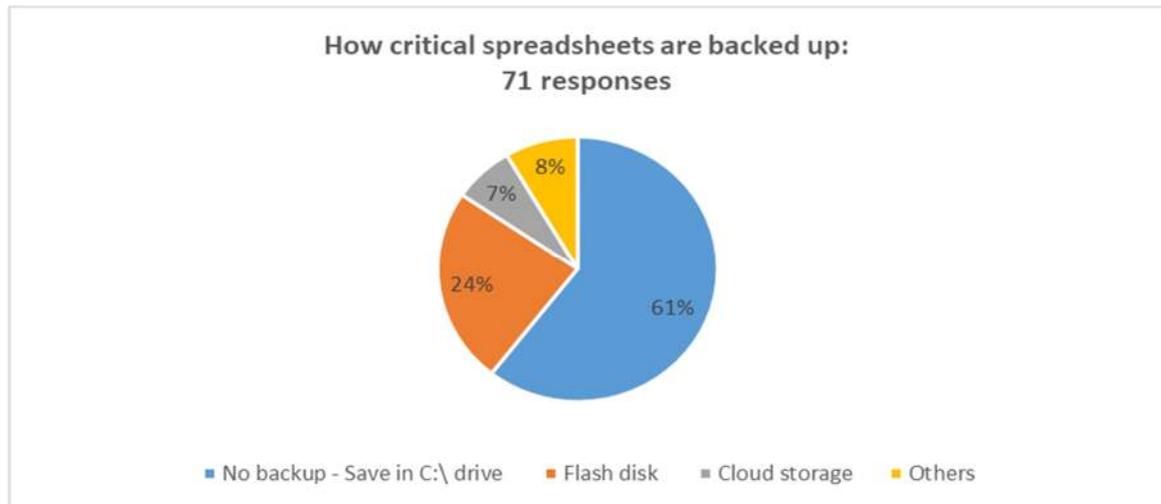

The poor backup strategies depicted above potentially expose the organization to risks such as loss of critical data, spreading of malware through removable drives and abuse of safety-critical data in water reticulation plants. A high level of such risks can lead to a complete business shutdown in the event of a disaster simply because spreadsheets play critical business functions as was the case with the BNFL in 1999 (Thorne, 2013).

In terms of cataloguing, Harrods (1990), defines the term as the process of compiling a list of documents based on set rules for easy retrieval from the archive. In this organisation, spreadsheets are critical information assets containing vital records for customers, employees, and operational processes. This calls for the need to organize spreadsheets for easy access by the users and other stakeholders.

| *Table 11: Information recorded when cataloging* | *Responses* | *Percentage* |
|---|---|---|
| *I do not catalogue* | *52* | *73%* |
| *Department* | *0* | *0%* |
| *Name of creator* | *6* | *8%* |
| *Others* | *3* | *4%* |
| *Version number* | *0* | *0%* |
| *Time/Date* | *10* | *14%* |

The results presented in Table 11 show that over 70% of spreadsheets are not catalogued. The practice is however common amongst 27% of the sample who indicated either time and date or name of the developer are typical classification parameters. In addition, the picture in Table 12 is more depressing as the majority of the sample does not only keep archives but also they seem not to understand the concept as a whole. Responding to the question on the frequency of archiving spreadsheets, cumulatively 47% indicated "strongly disagree" and "agree" whilst 49% indicted "neither agree nor disagree".

| *Table 12: Archiving is done regularly?* | *Responses* | *Percentage* |
|---|---|---|
| *Neither Agree nor Disagree* | *35* | *49%* |
| *Disagree* | *17* | *24%* |
| *Strongly Disagree* | *16* | *23%* |
| *Agree* | 3 | *4%* |
| *Strongly Agree* | 0 | *0%* |

Implications of this practice may include a high risk of losing vital information. As noted, there is a lack of adequate classification rules to facilitate cataloguing hence rendering even the 4% of spreadsheets that are archived as shown in table 12 are difficult to be fetched or accessed.

### 3.2.7 Staff Training

Under the training section, respondents were asked to specify any spreadsheet training that was made available whilst working with the organisation. As summarised below, about 70% of employees have never attended any kind of spreadsheet training. The remaining 31% attended some training, of which the majority indicated in-house trainings which tended to be elementary in nature according to remarks made by the respondents and the results presented in Table 13.

| *Table 13: Spreadsheet training made available* | *Responses* | *Percentage* |
|---|---|---|
| *None* | *49* | *69%* |
| *In-house training* | *12* | *17%* |
| *External training* | *7* | *8%* |
| *Others* | *3* | *4%* |
| *Several sessions and advanced topics* | *0* | *0%* |

The study further discovered that unavailability of trainers in Malawi and poor quality of training are major challenges. The results presented in Table 14 confirm. These findings conclude that people strongly believe that training scope available in the country tends to be too rudimentary for their daily spreadsheet activities and demands.

| *Table 14: Impediments to spreadsheet training* | *Responses* | *Percentage* |
|---|---|---|
| *Unavailability of local spreadsheet trainers* | *39* | *55%* |
| *Available training is of poor quality* | *19* | *27%* |
| *Inadequate support from management* | *9* | *13%* |
| *Lack of personal interest* | *4* | *6%* |

While the employees that have attended training account for only 30% according to the results presented in Table 13, it appears that the majority of that percentage attended in-house sessions covering spreadsheet basics. Despite that, results show that training sessions on advanced topics are scarce, 15% of the respondents indicated that they attended some training on advanced techniques, data analysis, and macros as displayed in Table 15.

| *Table 15: Topics covered in the training sessions* | *Responses* | *Percentage* |
|---|---|---|
| *None* | *49* | *65%* |
| *Basic techniques* | *11* | *15%* |
| *Data analysis* | *8* | *11%* |
| *Advanced techniques* | *4* | *5%* |
| *Macros* | *2* | *3%* |
| *Specialized add-ins and other tools* | *1* | *3%* |

Although the organization is faring badly as far as spreadsheet training is concerned, the level of dependence on spreadsheeting remains critical. When asked about prior experience, only 3% indicated "extensive experience" implying that 97% of users do not have the necessary skill set considering the criticality and extensive use of spreadsheets in the organisation. This situation would probably lead to a higher error rate than usual (Panko, 2002). The general position of survey results under this section shows a trend that resembles that of (Pemberton and Robson, 2000 and Baker et al, 2006) in the sense that advanced spreadsheet training is rare in many organisations hence the need for collaborative efforts to advocate for spreadsheet training that meets the needs of people in organizations.

### 3.2.8 Standards and Policies

Matters of standards and policies relating to information technology in Malawi are generally at the developing stage. The country adopted its first national ICT policy in 2014. While this policy tries to tackle some issues, especially the mainstreaming of technology in sectors of the economy, nothing is mentioned about spreadsheets and risks associated with their use. In the organisation under study, over 80% of the staff do not understand the concept of standards and policies, 11% said there are no standards at all whilst the remaining 3% hinted at the existence of verbal guidelines as summarised in figure 4.6. Similarly, results show that over 95% further reported that standards and policies are not followed whilst the rest of the sample indicated "neither agree nor disagree".

**Figure 6:** Spreadsheet Standards and Policies

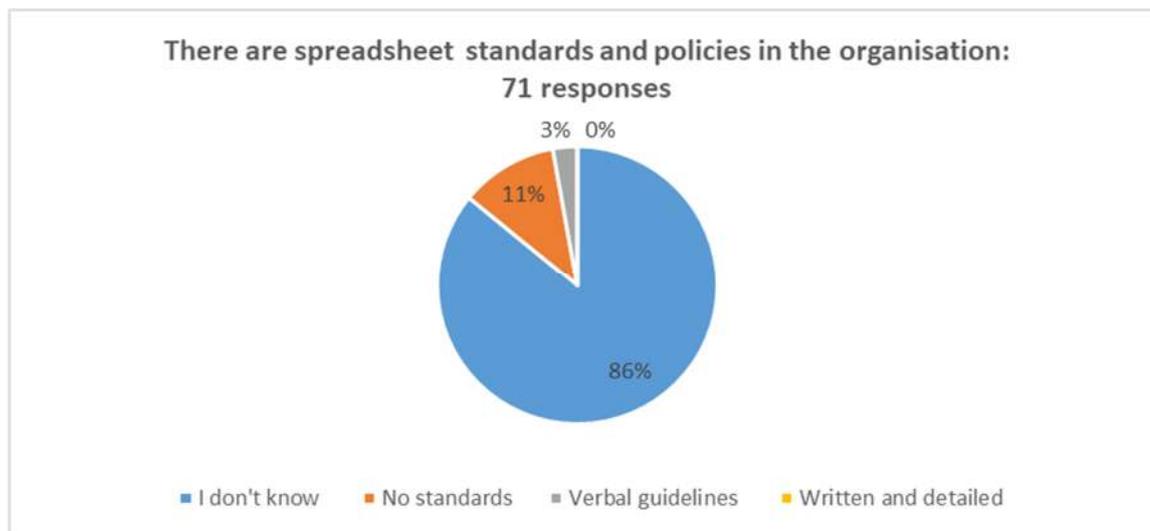

The current position explains why the organization was poorly rated based on survey results, suggesting the need to improve and adopt spreadsheet standards and policy implementations. According to (Thorne, 2013, Baker et al. 2006), well laid out standards help to define the procedures to be followed when working with spreadsheets in terms of spreadsheet life cycle phases.

| Table 16: Why standards and policies are not followed | Responses | Percentage |
|---|---|---|
| *I don't understand the standards* | *39* | *55%* |
| *No enforcement* | *33* | *46%* |
| *Others do not follow the standards* | 11 | *15%* |
| *They are irrelevant* | 0 | *0%* |
| *Too stringent* | 0 | *0%* |
| *No impediments* | 0 | *0%* |

Table 16 represent what people believe are some of the factors that contribute to non-adherence to the standards in the organisation. As the majority of the staff seem not to understand spreadsheet standards and policies, chances are very high that they may not even follow or use them if they were available. Therefore, the researcher suggests that awareness programs on spreadsheet standards and policies could best approach to improve spreadsheet governance worthy further exploration in the last chapter of this paper.

**Figure 7:** Spreadsheet Risks awareness

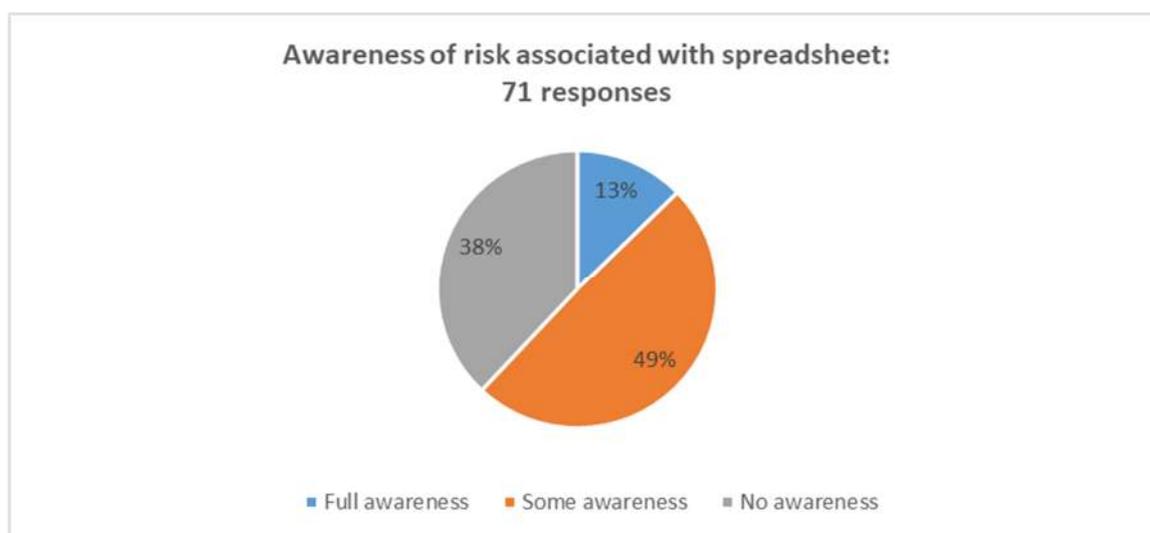

Regarding risk management activities, the organisation seems to be faring well compared to the rest of the research themes. The preliminary study discovered that the organization implemented a Risk Management Framework that focuses on the overall risk landscape in organisations. Therefore, it can be argued that some respondents did not really think of risk awareness in terms of spreadsheets considering the overall trend of responses to similar questions. The discussed phenomena are highlighted in figure 7. In table 17, 40% of employees are not sure about who manages spreadsheet risks, however, good proportions of the sample seem to be aware. The tabulations indicate that 27% of subjects think managing risks is a responsibility for developers, 20% think users are responsible and 9% believe that managers must lead in the risk management process.

| *Table 17 Who manages spreadsheet risks* | *Responses* | *Percentage* |
|---|---|---|
| *I don't know* | *31* | *41%* |
| *The Developer* | *20* | *27%* |
| *The User* | *15* | *20%* |
| *The Manager* | *9* | *12%* |
| *Insurance* | *0* | *0%* |

### 3.3 Results - Spreadsheet Audit

The researcher discovered several complex spreadsheet artefacts that are used in all three business units. However due to time constraints, the researcher selected a model that is largely used in the operations department to be audited using the XLTest tool. This particular model was selected based on its complexity and criticality to the organisation as attested by users that were interviewed. Furthermore, the study established that the organisation relies on this model to capture, track, monitor and analyse meter reading data collected from customer water meters in the field. This model was designed to enable validation of data entry completion before its contents are uploaded into the utility billing system. The model also helps to keep track of all newly installed meters on the ground by matching records in the billing system. Several parameters were implemented for tracking data depending on the needs at a particular time. The model also contributes some of the critical decisions that are made in the organisation based on trends displayed by this model such as monthly water production volumes, sales and cash collection. According to Chambers and Hamill (2008), in an environment where very large transactions are involved, massive losses can be orchestrated by even a very simple spreadsheet. Therefore, this complex spreadsheet model causes potential risks such as skipping customers during the door to door capturing of meter readings; abusing critical meter reading data by the Meter Readers; poor data integrity that can lead to senior management to make decisions based on flawed data.

#### 3.3.1 Spreadsheet Auditing Applications

In this study, XLTest was used to audit a spreadsheet sample in order to illustrate potential risks that can cripple business operations. XLTest is a powerful spreadsheet auditing application that established risky practices associated with spreadsheets in the organisation. The application offers several functionalities for performing the analysis. The detailed documentation reveals all the non-obvious content of the spreadsheet such as revealing hidden rows, columns, and sheets. A feature is provided for comprehensive visualization of the structure and content of a spreadsheet such as inconsistent formulas and data which stand out for attention. Another interesting aspect of XLTest is the fact that it offers a detailed error checking mechanism that makes it easy to find and fix errors far more quickly than with tedious cell-by-cell inspection. The test case maintenance and documentation make it easy to prove regression testing whilst the utilities feature to provide more approaches to test and profiling VBA performance. It also important to mention that several spreadsheeting auditing tools are current in use. For example, Grossman and Ozluk (2010), recommend Operis Analysis Kit (OAK) developed by Operis who are the largest provider of spreadsheet training on financial models (Colver, 2010).

#### 3.3.2 The Audit Results

XLTest spreadsheet auditing software has diverse analytics and audit functionalities. A summary of the batch audit log of the spreadsheet described in 3.3 reveals that the model contained 10 worksheets and listed 20 items involving the operations executed pertaining to distinct formula, data validation formula, data type and security mechanisms applied onto the selected model. The log also revealed that 5 hidden sheets, 8 protected sheets 11 window settings and 12 hidden columns were reset by the tool in order for the audit to be conducted. Bregar (2008) discusses several spreadsheet formula complexities and proposes that the metrics require further validation. One important complexity element is denoted by the nesting intensities of formula operators and operands. The sample audited contained 61 distinct formulas, accordingly, the formula complexity score in the audit results range from 1 to 97 with the latter being the most complex. It was also observed that some formulas were repeated separately up to 1186 times. High prevalence of hardcoded constants in formulas as revealed by the audit echoes huge risk probability the model possesses in the event of future reuse. By using cell error rate (CER) metrics as discussed by Panko (2008), undetected errors may be existing in about 1990 cells. It was also discovered that 172 constants were hard-wired in formulas and they were used in 3527 cells. Blayney, (2008), alleges that hardcoded values into spreadsheet formulae brought about similar challenges as in conventional programming. In this case, changing the constant value would require editing 172 formulas appearing in all the 3527 cells of the workbook. Clearly, this practice is bad and increases the catastrophic risk of making both logical and mechanical errors. Blayney (2008) recommends assigning a constant value to a variable name as the most appropriate practice.

The XLTest results further reveals many ridiculous long nested IF statements suggesting that the developers of this particular critical model do not know how to use built-in functionalities like VLOOKUP and PivotTable that would be simpler in many cases to reduce chances of missing invalid codes. Audit results exposed 281 instances of nested IFs of which 24 had errors in one model, meaning the risk exposure due to error is much higher if all the spreadsheets are aggregated. Zhang et al. (2018) discuss that nested-IF statements bring about readability and high cognitive challenges to spreadsheet users. They also lead to errors during the reuse or maintenance phase. These challenges point to computerised systems deficiencies in the organisation, forcing staff to adopt spreadsheets despite the fact that most of them do not have the expertise in spreadsheets to handle and realize problems with nested-IF expressions. The spreadsheet also contains defined names with #REF! values which look like the leftover of copy/paste from source workbooks. Additionally, there is the use of custom number format that forces a third decimal place to a zero: 0.00"0". The developer also used two-letter codes instead of validating data with a dropdown. With XLTest, lack of documentation makes improvements suggestions on the spreadsheet to be rather difficult other than the obvious formula simplification. The presence of hidden sheets and availability of user instructions also means that it is not obvious how it is updated and reported on.

OAK was used to compute a risk score by multiplying the risk weighting coefficient with the sum of all distinct formulas that were used in the workbook, accompanied by the total count of occurrences of selected formula elements. The resultant figure was 571, which was the overall risk score. OAK has other functionalities for listing all discrete formulas used along with narrations and complexity degree. The tool also flagged all formulas that were returning an Excel error message like #REF #ERROR as "Error formulae". These, however, might not be real errors and there was a need for them to be looked at individually.

Finally, the software also compiled and listed the details of all constants that are being referred to in a particular worksheet. These functionalities are helpful to track changes in different versions. However, due to limitations like time constraints, the researcher was unable to explore all the functionalities of the software tool in deeper detail. These audit results are just the tip of an iceberg. The risk analysis presented in this paper is a very small fraction since only one workbook was analyzed. And the fact that this spreadsheet model is duplicated in over 21 locations is potentially risky. Therefore, these results support the need for comprehensive research on spreadsheet audits considering that several other critical spreadsheets are currently in circulation with both internal or external and external stakeholders.

### 3.4 Summary of risks in the organisation

Based on the survey data, the organisation is engaging itself at considerable risk in different ways including the risk of making invisible errors in spreadsheets that propagate to strategic decision making. There are also risks of circulating flawed data to third parties and project financiers via data entry or erroneous analysis in spreadsheets. There is a risk of fraudulent activities going unnoticed through lack of ownership and standards governing spreadsheets, for instance, meter readings can be altered at the data entry stage by Sales Representatives for the spreadsheet discussed in 3.3. From finance, there is the risk that financial planning, forecasting and tracking could be compromised resulting in internal failure to plan and budget, there could be dire financial consequences as a direct result. For the operations section which is more production-focused, risks exist in the mishandling of data, errors in critical customer data and errors in reporting to external bodies which could put the institution in legal and financial jeopardy.

### 3.5 Criticality

From the data available, spreadsheets are very critical in the business operations of the organisation on a number of levels. All pre-billing activities, as well as financial planning and reporting, are conducted using spreadsheets as evidenced by responses to the survey. By using criticality tests developed by Chambers and Hamill (2008), McGeady and McGouran (2008) or Thorne and Shubbak (2016), volumes of spreadsheet applications would be deemed critical and in need of management. The data obtained in this survey points to numerous areas that need immediate and focussed consideration to reduce the possibility of catastrophes arising due to spreadsheet use.

### 3.6 Recommendations

The fact that IT development is still in the early stage in the organisation as demonstrated by the survey supports the presence of ample room for improvement. A wide range of both short and long term approaches can be chosen to reduce potential risks associated with the use of spreadsheets.

### 3.6.1 Key Risk Indicator (KRI) metrics

According to Thorne and Shubbak (2015), the risk calculator is one tool that can be used to generate a relative and absolute risk score for each spreadsheet used in the organisation. Ideally, the first step is to assemble active critical spreadsheet applications in order to identify applications with the highest risk requiring immediate attention. The Key Risk Indicator (KPI) tool for spreadsheets will be helpful to the organisation in the process of addressing the existing issues by setting priorities for mitigation activities aligned with the nature of the risk posed by a particular spreadsheet. Furthermore, mandatory actions are needed regarding the minimum controls such as version control, access control, change control, business continuity measures, documentation, and testing as discussed by Chambers and Hamill (2008). The management should ensure that every critical spreadsheet used in the organisation conforms to this requirement.

### 3.6.2 Spreadsheet Best Practices Policy

The organisation should promptly develop and enforce EUC Policy to provide clear guidelines and ensure that employees are accountable for their activities involving the use of spreadsheets. This approach will help to drastically reduce the risk of spreadsheet abuse by users as they will be aware of consequences to be borne for every action taken, hence protecting the organization's financial and information resources. The policy should further provide guidelines for spreadsheet creation, documentation and testing processes with regards to their characteristics and demands. The organisation should also consider the adoption of structured spreadsheet modelling and implementation (SSMI) methodology during the design and development phase. Mireault (2016) asserts that SSMI methodology is ideal for both experienced spreadsheet developers and non-computing professionals. The prevalence of artefacts that have become permanent assets for organizations greatly supports the need for comprehensive documentation within the spreadsheet as recommended by Pryor (2004) and O'Beirne (2005). In addition, the organisation needs to improve spreadsheet testing approaches. Diverse strategies are advocated by Panko (2006) including the "Fagan method'' tailored for spreadsheets. Panko hints that code inspection and auditing are the most effective approaches for detecting and correcting errors. Simple peer review and management review of spreadsheets offer cheap but effective approaches to augment risk mitigation measures. In short, the organisation needs to develop strategies for ensuring that the development of critical IT applications is carried out by trained IT developers while acknowledging critical applications developed by end users (McGeady and McGouran, 2009).

### 3.6.3 Identify and provide suitable training

Survey results discussed in section 3.2.2 show that spreadsheets are used for different purposes across the organisation. The results presented in section 3.2.7 further confirm that spreadsheet training is not common in the organisation, a common trend as depicted by other earlier studies revealing that spreadsheet training usually covers basics (Baker et al., 2006). The current situation justifies the need to recommend the introduction of specialized training for employees in the organisation. The organization's training and development section should, therefore, firstly categorize users according to their needs within the organization in order to identify appropriate sessions befitting the diverse users' categories in the organization. In this case, however, general training covering the basics of spreadsheet development techniques, documentation, testing, and other best practices is deemed as a mandatory requirement for all typical spreadsheet users.

### 3.6.4 Improving the Ergonomics

The billing process is one of the critical financial and operational processes that take place on a monthly basis in the organisation. This involves manual capturing of meter readings for over 174,000 customers into a spreadsheet that is duplicated across over 21 offices. While most interviewees said the task was simple to accomplish, but clearly this tends to be slow and prone to errors and extremely repetitive, hence vulnerable to lapses in attention and errors (Thorne, 2013). As discussed by Panko (2005), the error rate of data entry is approximately 0.5% for repetitive 'simple' tasks such as capturing a reading from a device and recording it on paper. The study further concluded that more logically complex undertakings lead to a higher error rate of about 5%. Therefore, it is likely that employees in the organisation certainly commit a higher base error rate under the current working conditions hence the need to mitigate the risk. The organisation should therefore consider implementation of automated data capturing techniques including the adoption of on-the-spot billing technologies, phasing out the use of paper-based metering reading by introducing user-friendly computer-based meter reading approaches like tablets and personal digital assistants to ensure that the efficiency of people does not deteriorate while working with spreadsheets.

## 4.0    Conclusions

In summary, the approaches to using spreadsheets presented in this paper are typical of most organisations. There is a high dependence on spreadsheets, lack of awareness over the prevalence of errors and the risks posed by spreadsheet use. Staff developing spreadsheet models are not trained and generally have 'self-taught' experience, they do not plan, develop with a methodology or test their models. There are no standards or policies whatsoever governing the use of spreadsheet artefacts within the organisation. As a result, the organisation places itself at severe risk of a decision being made on flawed numbers, assumptions or models for which the consequences could be severe. This paper has emphasized those points and others in detail and provides a basis for mitigating risks associated with end-user computing spreadsheet use in the organisation.

**Appendix – Questionnaire**

**EXPLORING SPREADSHEET USE AND PRACTICES IN A TECHNOLOGICALLY CONSTRAINED SETTING. "CASE STUDY OF NORTHERN REGION WATER BOARD IN MALAWI"**

---

**SECTION A: SPREADSHEET USAGE**

1. Which software do you use as spreadsheet package?

    - ☐ Microsoft Excel
    - ☐ OpenOffice
    - ☐ Google Sheets
    - ☐ LibreOffice

2. What are the key purposes of your spreadsheets you use?

    - ☐ Maintaining list e.g. names and addresses
    - ☐ Tracking data, e.g. budgets, sales, inventories
    - ☐ Analysing data e.g. financial, operational
    - ☐ Determining trends and making projections
    - ☐ Evaluating alternatives

3. How frequent do you use spreadsheets?

    - ☐ Daily
    - ☐ Weekly
    - ☐ Monthly
    - ☐ Quarterly
    - ☐ Annually

4. How do rate the level of spreadsheet importance in your roles:

    - ☐ Unimportant
    - ☐ Moderately Important
    - ☐ Very important
    - ☐ Critical

5. Which techniques best describes your general application of spreadsheets?

    - ☐ Statistical analysis
    - ☐ Optimisation
    - ☐ Simulation
    - ☐ None of the above

## SECTION B: DESIGNING

6. Do you design spreadsheet workbooks from scratch?

   - [ ] Never
   - [ ] Usually
   - [ ] Sometimes
   - [ ] Always

7. Please select practices that portrays initial step while designing a spreadsheet for your role in your department:

   - [ ] Borrow a design from existing spreadsheet
   - [ ] Sketch on paper
   - [ ] Write the fundamental relational algebra
   - [ ] Capture data and formulas directly
   - [ ] Use of models created by co-workers

8. Spreadsheet are slip into discrete modules when creating:

   - [ ] Never
   - [ ] Sometimes
   - [ ] Usually
   - [ ] Always

9. All inputs are separated from the formulas in my spreadsheets:

   - [ ] Never
   - [ ] Sometimes
   - [ ] Usually
   - [ ] Always

10. Do you follow standard development methodology e.g. Structured Spreadsheet Modelling and Implementation (SSMI) during development?

    - [ ] Never
    - [ ] Sometimes
    - [ ] Usually
    - [ ] Always

11. What is the intended usage approach of spreadsheet models you create?

    - [ ] For personal use
    - [ ] Shared with 1 - 2 people
    - [ ] Used by several people
    - [ ] Often become permanent assets in NRWB

**SECTION C: TESTING AND DOCUMENTING**

12. Do you perform spreadsheet testing activities?

    ☐ Never
    ☐ Rarely
    ☐ Frequently
    ☐ Always

13. Which methods are used to test your spreadsheets?

    ☐ Test extreme cases
    ☐ Check cells using calculator
    ☐ Common sense
    ☐ Display formulae
    ☐ Individual formula examination
    ☐ Test performance for integrity
    ☐ Go To - Special
    ☐ Error checking option
    ☐ Formula auditing toolbar

14. Do you document spreadsheets when designing or modifying?

    ☐ Never
    ☐ Sometimes
    ☐ Usually
    ☐ Always

15. What techniques do employ when documenting your spreadsheets?

    ☐ Text in spreadsheets
    ☐ Cell comments
    ☐ Documentation sheet
    ☐ Separate document
    ☐ Other

16. Approximately what percentage of your time during design phase is spent on your spreadsheet?

    ☐ 1 - 5%
    ☐ 6 - 15%
    ☐ Over 15

## SECTION D: SHARING AND MODIFYING

17. In what manner do you share spreadsheets with other users?

    ☐ I rarely share any part of a spreadsheet
    ☐ I provide a summary of results
    ☐ I provide parts of the spreadsheets
    ☐ I share the entire model

18. How frequent do you share spreadsheets with other users?

    ☐ Daily
    ☐ Weekly
    ☐ Monthly
    ☐ Annually

19. Who makes improvements or modifications to the spreadsheets used in NRWB?

    ☐ Original developer
    ☐ New developer
    ☐ Any users

## SECTION E: PROCEDURES OF ARCHIVING SPREADSHEETS

20. What approaches are employed for performing backups of critical spreadsheets in NRWB?

    ☐ No backup - save to C: rive
    ☐ Removable flash drives
    ☐ SharePoint or backup server
    ☐ Cloud storage e.g. Google drive
    ☐ Others

21. What information is recorded on your spreadsheets when cataloging archived spreadsheets?

    ☐ I do not catalogue
    ☐ Version number
    ☐ Time
    ☐ Name of creator/ developer
    ☐ Department
    ☐ Others

22. All critical spreadsheets that are used in NRWB are always archived for safe keeping:

| Strongly Agree | Agree | Neither Agree nor Disagree | Disagree | Strongly Disagree |
|---|---|---|---|---|
|  |  |  |  |  |

## SECTION F: SPREADSHEET TRAINING

23. What spreadsheet training have been made available?

    - ☐ None
    - ☐ In-house training
    - ☐ Training by an external party
    - ☐ One basic session is available
    - ☐ Numerous sessions including, advanced topics
    - ☐ Specialist trainer dedicated to assist developers and user

24. Which of the following topics were covered in the training programs offered to you?

    - ☐ A basic spreadsheet technique e.g. copy and paste
    - ☐ Advanced techniques e.g. built-in functions
    - ☐ Data analysis e.g. pivot tables
    - ☐ Use of specialised add-ins and other tools
    - ☐ Macros

25. What do thing are the biggest challenges to spreadsheet training sponsored by NRWB?

    - ☐ High cost due to unavailability of local trainers
    - ☐ Lack of personal interest
    - ☐ Available trainings are of poor quality
    - ☐ No management support
    - ☐ I prefer not to answer

## SECTION G: STANDARDS, POLICIES AND RISKS

26. Are there standards and policies relating to spreadsheets in NRWB or Malawi?

    - ☐ No standards
    - ☐ Verbal guidelines
    - ☐ Basic written standards
    - ☐ Written and detailed
    - ☐ I don't know

27. The spreadsheets standards and policies are followed in NRWB:

| Strongly Agree | Agree | Neither Agree nor Disagree | Disagree | Strongly Disagree |
|---|---|---|---|---|
|  |  |  |  |  |

28. What challenges affect following of the standards and policies relating spreadsheets?

- [ ] No challenges
- [ ] Too stringent
- [ ] Lack of knowledge in spreadsheets
- [ ] No enforcement
- [ ] Non-compliance to the standards by others
- [ ] I don't understand the standards
- [ ] I prefer not answer

29. What is your level of awareness on the potential risks that spreadsheets use poses to the organisation?

    - [ ] Full awareness
    - [ ] Some awareness
    - [ ] No awareness

30. Who is responsible for handling risks associated with the use of spreadsheets in NRWB?

    - [ ] The Developer
    - [ ] The User
    - [ ] The Manager
    - [ ] Insurance
    - [ ] I don't know

---

**SECTION H: DEMOGRAPHIC INFORMATION**

31. Please select your gender:

    - [ ] Male
    - [ ] Female

32. Please the range of your age in years:

    - [ ] 18 – 20
    - [ ] 31 – 40
    - [ ] 41 – 60
    - [ ] Over 60

33. What is your highest level of education?

    - [ ] High school
    - [ ] Undergraduate
    - [ ] Masters
    - [ ] PhD

34. Please select your position level or category:

☐ Casual
☐ Non- Manager
☐ Manager
☐ Executive

35. Choose department you are working in NRWB:

☐ Operations
☐ Finance
☐ Administration

**END OF QUESTIONNAIRE**